\documentclass[twocolumn,aps,nofootinbib]{revtex4}
\usepackage{amssymb}
\usepackage{amsmath}
\usepackage{epsfig}
\usepackage{tabularx}
\usepackage{inputenc}
\usepackage{url}
\usepackage{breqn}
\usepackage{hyperref}
\usepackage{footmisc}
\usepackage{titlesec}
\usepackage{fancyhdr}
\usepackage{graphicx}
\usepackage{caption}
\usepackage{slashed}
\usepackage{wrapfig}
\usepackage{color}
\graphicspath{{latex_images/}}

\begin{document}
\begin{titlepage}
\begin{center}
\hfill MI-TH-1760
\end{center}
\title{Bottom-quark Fusion Processes at the LHC for Probing $Z^{\prime}$ Models and B-meson Decay Anomalies}
\vspace{1.0cm}
\author{\textbf{Mohammad Abdullah, Mykhailo Dalchenko, Bhaskar Dutta, Ricardo Eusebi, Peisi Huang, Teruki Kamon, Denis Rathjens, Adrian Thompson} \\
\vspace{1.0cm}
\normalsize\emph{Mitchell Institute for Fundamental Physics and Astronomy, Department of Physics  and Astronomy, Texas A$\&$M University,
College Station, TX 77843}
\vspace{1.5cm}
}
\begin{abstract}
We investigate models of a heavy neutral gauge boson $Z^{\prime}$ coupling mostly to third generation quarks and second generation leptons. 
In this scenario, bottom quarks arising from gluon splitting can fuse into $Z^{\prime}$ allowing the LHC to probe it. 
In the generic framework presented, anomalies in $B$-meson decays reported by the LHCb experiment imply a flavor-violating $bs$ coupling of the featured $Z^{\prime}$ constraining the lowest possible production cross section. 
A novel approach searching for a $Z^{\prime}(\rightarrow\mu\mu)$ in association with at least one bottom-tagged jet can probe regions of model parameter space existing analyses are not sensitive to.
\end{abstract}

\maketitle
\end{titlepage}
After the Higgs boson discovery~\cite{Chatrchyan:2012xdj,Aad:2012tfa}, the major challenge for the LHC is to find new physics beyond the standard model (SM). 
Some intriguing excesses may hint at the presence of new physics. 
For instance, LHCb has reported an anomaly in the angular distribution of $B\rightarrow K^{*} \mu ^+ \mu^-$~\cite{Aaij:2013qta, Aaij:2015oid} (A similar anomaly has also been reported by the Belle Collaboration~\cite{Abdesselam:2016llu}).   
These measurements have poorly understood hadronic factors ~\cite{Ciuchini:2015qxb} which, however,  are less relevant for the LHCb measurements  of $R_K$ and $R_{K^*}$.  
Here $R_{K^*}$ is defined as $\frac{BR(B\rightarrow K^{*} \mu^+ \mu^-)} {BR(B\rightarrow K^{*} e^+ e^- )}$ and $R_K$  as $\frac{BR(B^+\rightarrow K^{+} \mu^+ \mu^-)} {BR(B^+\rightarrow K^{+} e^+ e^- )}$ and both show values lower than expected in the SM~\cite{Aaij:2014ora, Aaij:2017vbb}.  

Combining $R_K$ and $R_{K^*}$, the overall deviation  from the SM expection is at least at a level of 4$\sigma$~\cite{DAmico:2017mtc,Capdevila:2017bsm}. 
A massive $Z^{\prime}$ with a flavor changing $b s$ coupling, and a non-universal coupling to leptons could easily accommodate the $R_K$ and $R_{K^*}$ anomalies~\cite{Gauld:2013qja, Buras:2013qja, Buras:2013dea, Greljo:2015mma, Altmannshofer:2015mqa, Crivellin:2015mga, Crivellin:2015lwa,Niehoff:2015bfa, Falkowski:2015zwa, Allanach:2015gkd,Belanger:2015nma,Boucenna:2016qad, Chiang:2016qov, Bhattacharya:2016mcc, Crivellin:2016ejn,Alok:2017sui, Ko:2017lzd, Chiang:2017hlj, Ellis:2017nrp, Tang:2017gkz, Kamenik:2017tnu, Sala:2017ihs, DiChiara:2017cjq, Alonso:2017bff, Bonilla:2017lsq,Baek:2017sew,Alonso:2017uky,Bian:2017rpg}. 
Such a new gauge boson is featured in many beyond the SM theories where an extra $U(1)$ group has been proposed~\cite{London:1986dk,Cvetic:1996mf,Cvetic:1997ky,ArkaniHamed:2001nc,Hill:2002ap}.  
$Z^{\prime}$'s have been intensively searched for at the LHC, and the current limit is in the multi-TeV range~\cite{Khachatryan:2016zqb,ATLAS-CONF-2016-045}. 
While current $Z^{\prime}$ searches assume the $Z^{\prime}$ to couple to the first generation quarks and leptons, the current constraints on the  $Z^{\prime}$ from LHC searches and $B$ physics require the couplings to second and third generation fermions to be dominant.

We investigate a scenario that can satisfy the $B$-anomaly constraints in a dimuon final state.
We will show that when the $Z^{\prime}$ boson couples to $b$ quarks, it is possible to use the $b$'s arising from gluon splitting with a final state consisting of at least one $b$-jet and two muons. 
As these diagrams are very similar to vector-boson fusion diagrams, we will name them bottom-fermion fusion (BFF) in the following. 
A generic framework of a minimal extension to the SM which explains the $B$ anomalies is used to discuss the new search strategies in this BFF production process. 
This BFF production process allows us to interpret the inclusive dimuon searches in light of a $Z^{\prime}$ coupling to the third generation. 
We will show that the  presence of additional $b$ jets in the dimuon final state can be utilized to probe smaller $Z^{\prime}$ masses than the inclusive dimuon searches. 
This strategy can be utilized for any model where $Z^{\prime}$ couples to $b$ quarks irrespective of solutions to $B$ anomalies. We will show that the flavor violating $bs$ coupling produces a lower bound on the $Z'$ production cross-section.
We compare several signal hypotheses to the SM background showing the possibilities of probing the parameter space explaining the $B$ anomalies at the present and future LHC runs.


The new physics contribution to rare $B$ decays can be described by the following effective Lagrangian
\begin{equation}
\mathcal{L} \supset \frac{4\, G_F}{\sqrt{2}} \, V_{tb}\, V_{ts}^* \, \frac{e^2}{16\, \pi^2} \, C_9\, O_9 + \textrm{h.c.} .
\label{eq:lag0} 
\end{equation}
The effective operator $O_9$,
\begin{equation}
O_9  = (\,\bar{s}\,\gamma_{\mu}\, P_L b\,)\, (\, \bar{\mu}\,\gamma^{\mu}\,\mu\,),
\end{equation}
describes a four-fermion interaction, with a left-handed $b\rm{-}s$ current and a vector current for $\mu$.
To fit the current data~\cite{Altmannshofer:2017yso}, the new physics contribution to $C_9$ needs to be $-1.59^{+0.46}_{-0.56}$. 

Here, we consider a toy model by extending the SM by adding an extra $U(1)$ gauge group, which introduces a new gauge boson $Z^{\prime}$. 
With a flavor changing quark coupling and a non universal lepton coupling, it can generate a contribution to the desired effective operator. 
The minimal phenomenological Lagrangian is
\begin{equation}
\mathcal{L} \supset Z^{ \prime}_{\mu} \, [ \,g_{\mu} \,\bar {\mu} \,\gamma ^{\mu} \,\mu \,  + \, (\,  g_b\,  \delta_{bs}\, \bar{s}\,\gamma^{\mu}\, P_{L}\, b \, + \, \textrm{h.c.} \, )\, ]
\label{eq:LagMin}
\end{equation}
The contribution to the effective $O_9$ operator is 
\begin{equation}
\frac{e^2} {16 \pi^2} \, V_{ts}^*\, V_{tb} \, C_9 \, = \, - \frac{v^2} {2 {m^2_{Z^{\prime}}}} \,  g_b \, \delta_{bs}\, g_{\mu}.
\end{equation}
 which, using the central value of $C_9$, leads to the requirement  
 \begin{equation}
 g_b\, \delta_{bs} \,g_{\mu} (100\, \text{GeV}/m_{Z'})^2 \simeq 1.3 \times 10^{-5}
 \end{equation}

To evade the current bounds from the LEP and the LHC, we consider a scenario where the $U(1)$ charges of the fermions are flavor dependent as done in many studies \cite{Sierra:2015fma, Crivellin:2015mga, Niehoff:2015bfa, Boucenna:2016qad, Altmannshofer:2014cfa} to generate Eq.~\ref{eq:LagMin}. 
The $Z^{\prime}$~to~$bs$ coupling may, for instance, be generated from the mixing of vector-like quarks and leptons with their SM counterparts. 
In the lepton sector, the $Z^{\prime}$  needs to couple only to the muons. 
In order to preserve $SU(2)$ invariance, the $Z^{\prime}$ also couples to tops and muon neutrinos. 
We can write the following dominant terms in the Lagrangian in a model which contains Eq.~\ref{eq:lag0} and is allowed by all the existing constraints in order to address the anomalies:
\begin{align}
\label{eq:lag}
\mathcal{L} \supset Z^{ \prime \mu} \, [ &\,g_{\mu} \,\bar {\mu} \,\gamma ^{\mu} \,\mu + \,g_{\mu} \,\bar {\nu_{\mu}} \,\gamma ^{\mu} \, P_{L} \, \nu_{\mu} \,  \\ \nonumber
+&  g_b\,  \sum_{q = t,b} \bar{q}\,\gamma^{\mu}\, P_{L}\, q \,  + (\,  g_b\,  \delta_{bs}\, \bar{s}\,\gamma^{\mu}\, P_{L}\, b \, + \, \textrm{h.c.} \, )\,)]
\end{align}

The $Z^{\prime}$ mass is constrained to be less than 5.5 (10) TeV in the 1 (2) sigma range to explain the $B$ anomalies. 
It can be as light as $100$ GeV while still satisfying $B$ anomalies and other constraints~\cite{Altmannshofer:2014cfa}. 
As shown in Eq.~\ref{eq:lag}, the $Z^{\prime}$ does not significantly couple to first or second generations quarks thus weakening current limits on $Z^{\prime}$ production at the LHC. 
However, the $Z^{\prime}$ can be produced through its couplings to $b$ quarks originating either from sea quarks, or gluon splitting. 
Therefore, the $Z^{\prime}$ is associated either with two $b$-jets (both $b$ quarks from gluon splitting), one $b$-jet (one $b$ quark from each of gluon splitting and sea quarks), or no $b$-jet (both $b$ quarks from  sea quarks). 


The $Z^{\prime}$ will decay into pairs of $b$ quarks, muons, muon neutrinos, and, if kinematically allowed, top quarks. 
Therefore, the relevant final states at the LHC are dimuon or di-$b$ resonances.  The cross sections behave as follows:

\begin{align}
&\sigma(\, p p \rightarrow Z^{\prime} \, \rightarrow \mu\, \mu \,) \, \sim \frac{2\,g_b^2(1+k\delta^2_{bs})\, {g_\mu}^2} {6\, g_b^2\, + 3 g_{\mu}} \\
&\sigma(\, p p \rightarrow Z^{\prime} \, \rightarrow b\, \bar{b} \,) \, \sim \frac{3\,g_b^4(1+k\delta^2_{bs}) \,} {6\, g_b^2\, + 3 {g_{\mu}}^2}
\end{align} where $k$ contains the $s$-quark PDF effect since the production of $Z^{\prime}$ may occur through $bs$ fusion. When $\delta_{bs}$ goes to zero, the flavor conserving contribution dominates the production of $Z^{\prime}$. When $\delta_{bs}$ is large but still satisfies the $B$ anomalies (so smaller $g_b$) the flavor violating contribution dominates. 

Since the $Z^{\prime}$ is produced primarily through $b$ couplings and can decay into a pair of muons, bottom quarks, or tops, the searches for dimuon~\cite{Khachatryan:2016zqb,ATLAS-CONF-2016-045}, dijet~\cite{ATLAS-CONF-2016-030,Aaboud:2016nbq,Sirunyan:2016iap} or $t\bar{t}$~\cite{SanchezMartinez:2016nhj,Khachatryan:2015sma} resonances are relavant. 
The reliance on bottom quarks for production in our scenario weakens the impact of existing searches  compared to scenarios utilizing production via first generation quarks. 
Dijet and $t\bar{t}$ constraints are inconsequential since the uncertainty in the $t\bar{t}$ cross section measurement is several pb and the current 8 TeV constraint on the  resonance searches is  $\mathcal{O}(\textrm{pb})$ while the dimuon resonance searches produce relevant constraints.

In addition to direct searches for a  $Z^{\prime}$, its flavor changing coupling also generates a contribution to the $B_s - \bar{B}_s$ mixing, thereby changing the mass difference of $B_s$ mesons.
The current measurement of the deviation from the standard model is about $\Delta_{B_s}= 0.07 \pm0.09.$~\cite{Bona:2016bvr}. 
For a $Z^{\prime}$ of $\mathcal{O}(100)$ GeV, the $B_s - \bar{B}_s$ mixing is the dominant constraint~\cite{Altmannshofer:2014cfa} while other flavor constraints, such as muon $g-2$~\cite{Bennett:2004pv} and Br$(B\rightarrow  K{\bar{\nu}}\nu)$~\cite{Lees:2013kla, Lutz:2013ftz} are weak.

The measurement of neutrino trident production ~\cite{Altmannshofer:2014pba} places an upper bound on $g_{\mu}$ which, while too weak for our purpose, translates into a lower limit on the combination of $g_b\,\delta_{bs}$ that explains the $B$ anomalies.

Since the measurements of  $R_K$ and $R_{K^\ast}$ fix the combination ${g_b\delta_{bs}g_\mu}\over{m^2_{Z^{\prime}}}$, $Z'$ production through BFF dominates for large $g_b$ and, therefore, small $\delta_{bs}$ and $g_\mu$. For each value of $m_{Z'}$ we will fix $g_\mu$ such that $g_b\, \delta_{bs}$ has the maximum value allowed by $B_s$ mixing. 
When $\delta_{bs}$ becomes as large as about $0.6$, diagrams including $s$-quarks start dominating the production of  $Z^{\prime}$ and is not covered in the work. Note that the production through $b$ quarks alone can lead to dimuon events plus one or two $b$ jets, whereas the production via $b$ and $s$ quarks consists almost exclusively of 1$b$ + $\mu \mu$ final states. 

\begin{figure}[!tbh]
  \begin{center}
   \includegraphics[width=0.27\textwidth]{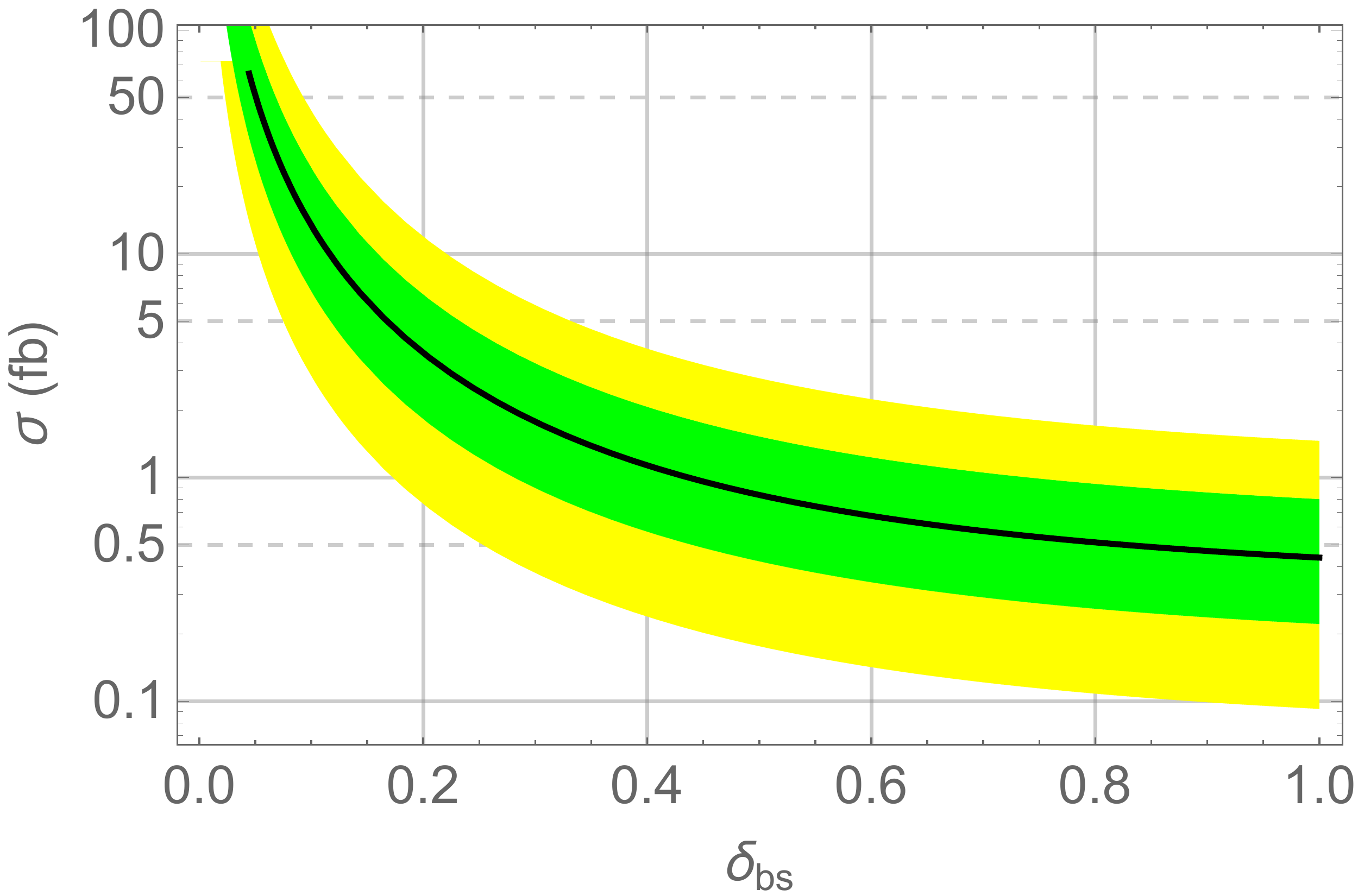}
    \caption{\label{fig:xsec0} Production cross section for a BFF dimuon resonance as a function of $\delta_{bs}$ for a 350 GeV $Z^{\prime}$ and $g_\mu=0.13$ which satisfies the LHCb constraints. 
    The central fit (black line) and the 1$\sigma$ (green shaded region) and 2$\sigma$ (yellow shaded region) contours of the $B$ anomalies are shown.}
    \end{center}
\end{figure}

Figure ~\ref{fig:xsec0} shows the range of production cross-sections for dimuon + $b$ or 2 $b$ final states for $m_{Z^\prime}=350$ GeV and $g_\mu=0.13$ as a function of $\delta_{bs}$ with central (black line), 1 sigma (green shade region) and 2 sigma (yellow shaded region) fits  of the $B$ anomalies. 
The allowed cross-section band has a smaller slope for  larger $\delta_{bs}$ due to the dominance of $g_b\delta_{bs}$ coupling initiated $Z^{\prime}$ production, whereas in the smaller $\delta_{bs}$ region, the $Z^{\prime}$ production is dominantly governed by the flavor conserving $g_b$ term which decreases as $\delta_{bs}$ increases. 
For particular masses, the central fit (1$\sigma$ range) minimum cross sections are 0.2(0.12) fb for $m_{Z^\prime}=500$ GeV, 0.6(0.2) fb for $m_{Z^\prime}=350$ GeV, and 1.2(0.8) fb for $m_{Z^\prime}=200$ GeV.
The BFF production of the dimuon final state allows us to rule  out a large region of parameter space. 
Existing constraints are weak for $m_{Z^{\prime}}\leq 500$ GeV due to the large SM background contributions in that region. 
For this allowed parameter space, we introduce a simplified search strategy for various mass points searching for  $Z^{\prime}\to\mu\mu$ with at least 1$b$ jet  in this subsection.

For the following study of expected limits and selection requirements, we use MadGraph5 v.2.5.4~\cite{Alwall:2014hca} to generate signal and background samples. We use a modified version of the FeynRules model file for the Hidden Abelian Higgs Model~\cite{DuhrFeyn} as well as a model file of our own \cite{Christensen:2008py}\cite{Alloul:2013bka}.
Pythia~8.2~\cite{Sjostrand:2014zea} is used for parton showering and Delphes~3.4~\cite{deFavereau:2013fsa} for the detector simulation with a default CMS card. 
We consider pileup effects to be mostly mitigated in a realistic experimental analysis, thus we did not include any.
Electron and muon candidates are restricted to $|\eta| < 2.5$ and $< 2.4$, respectively.
Jets are required to have $p_{\rm T} > 30$ GeV. 
The jet pair in our selection is always comprised of the leading b-tagged jet together with the next-to-leading jet that is b-tagged, if possible. 
Only if no second $b$-tagged jet with $p_{T}>30$ GeV exists, the leading non-$b$-tagged jet in transverse momentum is chosen instead. 
A medium working point of the identification of $b$ quark jets in the Delphes package is used, where the $b$-tagging efficiency is varying with $p_{\rm T}$~($85\%$ at $p_{\rm T} >$ 30 GeV) and the corresponding misidentification probability for gluon- and light-flavored quark jets is in the range of $10-20\%$.

For a normal search for a heavy resonance decaying to opposite sign (OS) dimuons, the main backgrounds are Drell-Yan and $t\bar{t}$ events. 
Requiring OS dimuons and at least two jets with $p_{T}>30$~GeV, at least one of them passing a $b$-tag requirement, reduces Drell-Yan (+0,1,2 jets) process contributions to the search region by a factor of $\mathcal{O}$(100). 
The remaining background in the mass range beyond the $Z$-boson peak is the di-leptonic $t\bar{t}$ process that we further suppress by a set of three selection requirements:

\begin{enumerate}
\item Top mass bound  $M_{\mu b}$: We examine the muon-jet invariant mass in both exclusive permutations out of the dimuons and the two jets ($bb$ or $bj$). 
Of the two possible muon-jet parings, we choose the one with the smallest mass difference and require the heavier mass to be greater than 170 GeV.
\item Leptonic versus hadronic activity: The scalar sum of transverse momenta of the leading OS muon pair ($L_{\textrm{T}}$) must be larger than the scalar sum of transverse momenta of the leading bottom-tagged pair or bottom and non-tagged jet pair ($H_{\textrm{T}}$). 
\item Normalized missing transverse energy ($E_{\rm{T}}^{\rm{miss}}$): The ratio of $E_{\rm{T}}^{\rm{miss}}$ to dimuon mass ($M(\mu^{+}\mu^{-})$) is restricted to below 0.2 to reject events with real sources of $E_{\rm{T}}^{\rm{miss}}$. 
\end{enumerate}
While the top quark mass bound and the normalized $E_{\rm{T}}^{\rm{miss}}$ are expected to be useful for reducing di-leptonic $t\bar{t}$ contributions, the difference between $L_{\textrm{T}}$ and $H_{\textrm{T}}$ is specific to the BFF initial state. 
In contrast to forward-backward VBF production with its typical large invariant mass and rapidity gap selection on forward jets, BFF jets are usually centrally produced and, due to the gluon-splitting nature of their production, soft. 
This can be used to select BFF-produced heavy resonances in favor of many SM background scenarios that favor more even distributions of transverse momenta without requiring the high momentum thresholds other initial states like boosted object searches necessitate. 
In addition, it is possible to use more stringent requirements on $H_{\textrm{T}}-L_{\textrm{T}}$ to generate even background-free selections for heavier resonance scenarios. The $m_{T2}$ variable \cite{Lester:1999tx} has also been tested and found not to significantly improve on the other three selection requirements. 
As the best performance of inclusive dimuon resonance searches moves to higher masses as expected background contributions rise with increasing integrated luminosity, we expect more stringent selections like BFF to become competitive in terms of exclusion power for an increasing range of masses.

\begin{figure}[!tbh]{
\includegraphics[width=.45\textwidth]{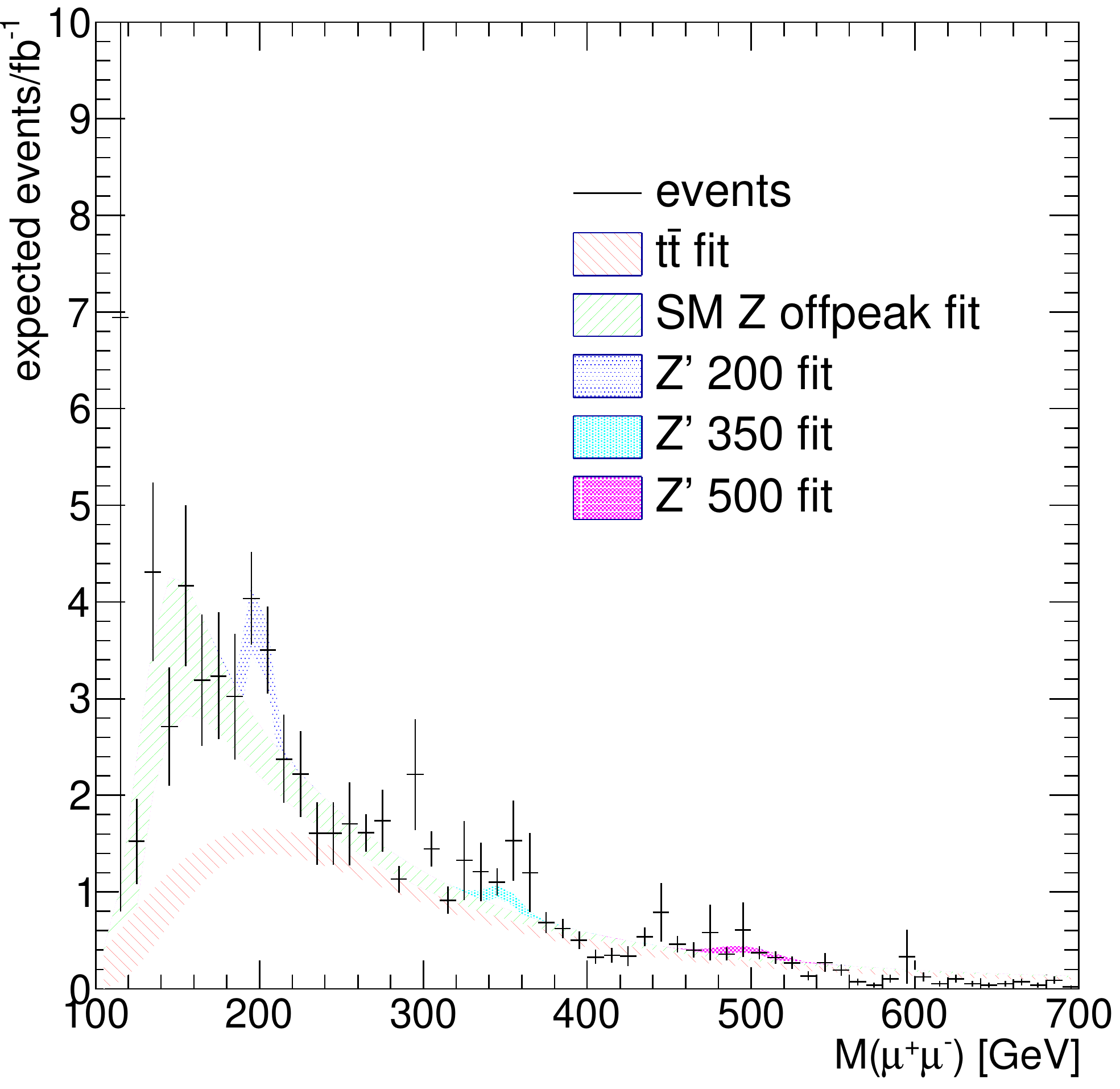}
\caption{Opposite sign dimuon invariant mass distribution for selected simulated events including the shape fits for background and signal contributions used to generate expected limits. 
Simulated data shows statistical uncertainties due to event weights only, uncertainty bands for the fits show the one sigma uncertainties of varying all fit parameters.}
\label{fig:mass}
}
\end{figure}

Table \ref{tab:selection} contains the efficiencies of the aforementioned selection requirements on di-leptonic $t\bar{t}$, SM $Z$ and three different mass scenarios for the $Z^{\prime}$ model.
\begin{table}[htbp]
\begin{tabular}{|l|r|r|r|r|}
\hline
 & \multicolumn{1}{l|}{preselection} & \multicolumn{1}{l|}{$M_{\mu b}$ } & \multicolumn{1}{l|}{$H_{\textrm{T}}-L_{\textrm{T}}$} & \multicolumn{1}{l|}{$E_{\rm{T}}^{\rm{miss}}$/$M(\mu^{+}\mu^{-})$}  \\ \hline
$t\bar{t}$ & 8\%   & 17\% & 26\% & 27\% \\ \hline
SM $Z$       & 0.2\% & 41\% & 32\% & 54\% \\ \hline
$Z^\prime$ 200     & 7\%  & 66\% & 76\% & 89\% \\ \hline
$Z^\prime$ 350     & 10\%  & 87\% & 90\% & 96\% \\ \hline
$Z^\prime$ 500     & 13\%  & 94\% & 94\% & 98\% \\ \hline
\end{tabular}
\caption{Efficiency of selection requirements for a simplified search for three different mass points assuming $\delta_{bs}=0$ with a dimuon $t\bar{t}$ background. 
The requirements are applied successively from left to right. 
Each entry indicates the individual requirement's efficiency after applying all other selections in columns to its left. 
The total efficiency of a background is the multiplication of all entries in a given row.}
\label{tab:selection}
\end{table}
The signal preselection is a function of $\delta_{bs}$ and $Z^\prime$ mass, as higher masses increase the hardness of associated jets and the centrality of events while higher values of $\delta_{bs}$ decrease the overall proportion of associated bottom jets compared to the total production cross section. 
We fit the dependence upon $\delta_{bs}$ with a linear fit for each mass point by generating several differently $\delta_{bs}$-valued samples with constant $g_{b}$. 
Then, we fit the resulting absolute values of slopes and intercepts versus $Z^\prime$ mass with a logarithmic fit each to determine a function describing the signal acceptance $A$ over the complete parameter space:
\begin{equation}
 A(m_{Z^\prime},\delta_{bs})=(0.063-0.026\delta_{bs})\textrm{ln}(\frac{m_{Z^\prime}}{\textrm{GeV}})-0.268+0.11\delta_{bs}
\end{equation}
Applying this selection yields Fig.\ref{fig:mass}. 
We use $g_{\mu}\sim$1 to calculate the $Z^{\prime}$ decay width to make sure our bound is valid for such high values of couplings. The values dictated by the $B$ anomalies are much smaller and would lead to a narrower width and hence a larger significance.    

\begin{figure*}[!t]
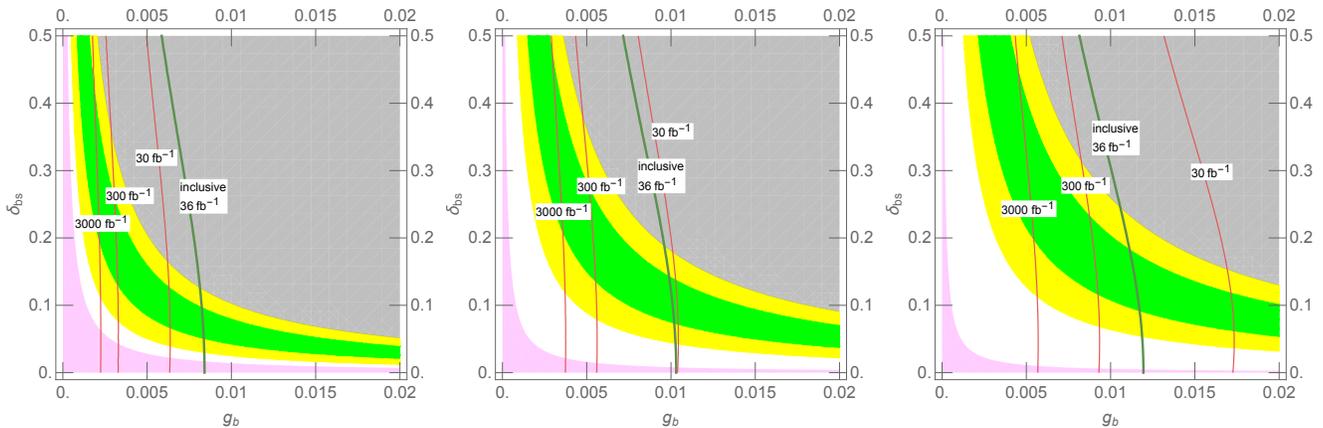

  \begin{center}
 \includegraphics[width=0.32\textwidth]{200GeV.pdf} 
 \includegraphics[width=0.32\textwidth]{350GeV.pdf} 
    \includegraphics[width=0.32\textwidth]{500GeV.pdf}
    \caption{\label{fig:xsec} The current and future expected LHC limits for various luminosities for three different $Z^{\prime}$ masses: 200, 350 GeV  and 500 TeV (left to right). 
    Green lines refer to the current reach of an inclusive dimuon search while red lines show the expected power of a 2(1)$b$ + dimuon search. 
    The yellow and green shaded regions correspond to 2$\sigma$ and 1$\sigma$ bands of the best fit to the $B$ anomalies for the chosen values of $g_\mu$ (0.08, 0.14 and 0.20 respectively). The grey shaded area is ruled out by the $B_s - \bar{B}_s$ mixing constraint. In the pink region, the required $g_\mu$ to fit the $B$ anomalies is ruled out by neutrino trident production.}
    \end{center}
\end{figure*}

Utilizing this new search strategy, we show the LHC reach for 200, 350, and 500 TeV $Z^{\prime}$ masses in the $\delta_{bs}-g_b$ parameter space in Fig.~\ref{fig:xsec} along with constraints from $B_s - \bar{B}_s$ mixing ~\cite{Altmannshofer:2014cfa} and trident production ~\cite{Altmannshofer:2014pba}. The 1$\sigma$ and 2$\sigma$ contours of the best $B$ anomalies fits are shown for the smallest $g_\mu$ values satisfying the mixing limits. The values of $g_\mu$ are 0.08, 0.14, and 0.20 for $M_Z^{\prime}$ masses 200 GeV,  350 GeV, and 500 GeV respectively. The 95\% exclusion limits for 30, 300 and 3000 fb$^{-1}$ of LHC integrated luminosity using 2(1)$b$ + dimuon final states are contrasted with the current and projected inclusive dimuon search limits \cite{Khachatryan:2016zqb}\cite{Khachatryan:2014fba}. Also shown are the regions excluded by neutrino trident production as explanations of $B$ anomalies. Note that these latter limits are on $g_\mu$ which has no bearing $g_b \delta_{bs}$ if we do not require a fit to $B$ anomalies. 

For $m_{Z^\prime}$=200 GeV, there is no current (or future) sensitivity from the inclusive dimuon search as the SM backgrounds render it insensitive in this region. 
However, a 2(1)$b$ + dimuon search as proposed in this work can probe a wider region of more background-ridden parameter space. 
Increasing $m_{Z^{\prime}}$ improves on the relative reach of the inclusive dimuon search due to reduces SM background expectations. 
At $m_{Z^{\prime}}=1~$TeV, the current search limit does not rule out any parameter space although increasing integrated luminosities should facilitate large improvements. 
The projected inclusive dimuon resonance search will be able to probe some parameter space up to $m_{Z^\prime}\sim$ 3 TeV, where the cross section is too small even for 3000 fb$^{-1}$.


In summary, we pointed out that the fusion of $b$ quarks from gluon splittings and sea-quark distributions at the LHC is vital for testing heavy $Z^{\prime}$ models where the $Z^{\prime}$ boson preferredly couples to quarks in the third generations. If such models are used to explain the $B$ anomalies, 
we show that there is a lower limit on such production processes that arises from the flavor-violating $b s Z^{\prime}$ coupling.
Producing such a $Z^{\prime}$ in a  final state is expected in association with one or two $b$ jets. 
The presence of the  resonance due to the BFF initiated processes allows us to probe such models in the inclusive  searches. 
Furthermore, the presence of additional $b$ jets along with kinematical requirements on them is found to be very effective in reducing SM backgrounds in background-dominated search regions (e.g., $\leq$ 500 GeV for the 13 TeV LHC). 
The prospects for testing the entire parameter space of such models  for some $Z^{\prime}$ masses appear to be excellent in the existing and upcoming LHC program.

MA, BD, RE, TK, DR and PH are supported in part by the DOE grant DE-SC0010813. MD, PH, and AT thank the Mitchell Institute for Fundamental Physics and Astronomy for support. 
TK is also supported in part by Qatar National Research Fund under project NPRP 9-328-1-066.
We would like to thank A. Datta, T. Ghosh, Xiao-Gang He, M. L. Mangano, J. Ruderman and J. Walker for useful discussions and comments.

\bibliographystyle{utphys}
\bibliography{zprime}
\end{document}